\documentclass[aps,osajnl,showpacs,superscriptaddress,10pt]{revtex4-2}

\usepackage{dcolumn}% Align table columns on decimal point
\usepackage{bm}% bold math
\usepackage{multirow,array}
\usepackage{graphicx}
\usepackage{amsmath}
\usepackage{float}
\usepackage{empheq}
\usepackage{adjustbox} % In preamble
\usepackage{booktabs}
\usepackage{multirow}
\usepackage[table,xcdraw]{xcolor}
\usepackage{url} % For proper URL handling
\usepackage[colorlinks=true, linkcolor=blue, citecolor=blue, urlcolor=red, filecolor=blue]{hyperref}
\usepackage{listings}
\lstdefinestyle{mathematicastyle}{
  language=Mathematica,
  basicstyle=\small\ttfamily,
  keywordstyle=\color{purple},
  commentstyle=\color{blue!50!black},
  stringstyle=\color{blue},
  showstringspaces=false,
  breaklines=true
    morekeywords={int,char,double,if,else,while,for,cout,cin, endl, open, ofstream,  chrono, steady_clock, now, Do, If, Abs, Print, Table, Solve },
  showstringspaces=false,
  breaklines=true
  columns=fullflexible
}

\usepackage{listings}
\usepackage{xcolor}
\lstdefinestyle{mystyle}{
  language=C++,
  basicstyle=\small\ttfamily,
  keywordstyle=\color{blue},
  commentstyle=\color{blue!50!black},
  stringstyle=\color{blue},
  morekeywords={int,char,double,if,else,while,for,cout,cin, endl, open, ofstream,  chrono, steady_clock, now, Do, If, Abs, Print, Table, Solve },
  showstringspaces=false,
  breaklines=true
  columns=fullflexible
}

\begin{document}

\title{A Search Method for Hirota Bilinear Systems of Nonlinear Evolution Equations}

\author{I. Albazlamit} 
\affiliation{Department of Physics at United Arab Emirates University, P.O. Box 15551, Al-Ain, United Arab Emirates}

\author{L. Al Sakkaf}
\affiliation{College of Engineering, Abu Dhabi University, P.O. Box 59911, Al-Ain, United Arab Emirates}

\author{U. Al Khawaja$^\ast$}
\affiliation{Department of Physics at United Arab Emirates University, P.O. Box 15551, Al-Ain, United Arab Emirates}
\affiliation{Department of Physics at The University of Jordan, Amman 11942, JORDAN}
\email{Corresponding author: u.alkhawaja@ju.edu.jo}

\begin{abstract}
We present a systematic search method for finding Hirota bilinear systems of nonlinear evolution equations, with emphasis on the nonlinear Schr\"odinger equation (NLSE). Using a known exact solution, couplings between the different terms of the differential equation are identified, which are then used to derive the bilinear system. We show that a nonlinear evolution equation may have many solution-dependent Hirota bilinear systems.
Nonetheless, all solution members of a certain solution class are associated with a single bilinear system. This has been demonstrated for the
known solutions of the NLSE. For instance, all solution members of the $N$-bright soliton solutions class lead to the same bilinear system. Similarly, the class of $N$-dark soliton solutions and the class of breathers solutions are associated with their own distinctive bilinear systems. Identifying the bilinear system of a known {\it seed} solution will
thus help in finding the rest of the solution members of the same class using the Hirota method. We apply our method to the nonintegrable NLSE with dual nonlinearity and external potential.

\end{abstract}

%\ocis{130.2790, 230.4555, 190.6135.}

\maketitle


\textbf{DOI:} https://doi.org/10.1016/j.nexres.2025.100705
\section{Introduction}
\label{introsec}

Finding new soliton solutions and studying their dynamics and characteristics play an important role in the scientific advancement of different fields due to their applications in quantum mechanics, plasma physics, hydrodynamics, fluid dynamics, optical fibers, and many others \cite{dhiman2024analyzing, kumar2024exploring, raza2022new, raza2023class}. 

The Hirota method is one of many different methods of studying integrable nonlinear evolution equations that have soliton solutions, such as, but not limited to, the inverse scattering transform (IST). However, the Hirota method applies to a broader class of equations than IST \cite{hirota2004direct,fokas2008unified,ablowitz1991solitons,rogers1982backlund,agrawal2000nonlinear,shabat1972exact, matveev1991darboux}, which is widely used in deriving soliton solutions \cite{hietarinta2013hirota, ye2011efficient, pashaev2022hirota, gurses2023method, hietarinta2007introduction}.

 The idea of the Hirota method is to transform the nonlinear evolution equation into a bilinear equation which can then be solved by the perturbation method \cite{hietarinta1997introduction,hietarinta2005hirota,ablowitz1986bilinear}. However, finding the Hirota bilinear system is the core technique used to search for soliton solutions. Basically, to find the bilinear system we write the transformed equation in terms of a special differential operator; the so-called Hirota D-operator, as defined below. 
The benefit of using the Hirota differential operator over the normal differential operator is that the nonlinear differential equation can be rewritten in terms of a bilinear system, which is easier to solve.

The appealing feature of the Hirota method is that it is relatively easier than other methods. 
In addition, existence of the Hirota bilinear system may be used as a criterion to establish the integrability of nonlinear evolution equations. The Hirota definition of integrability has proven to be analogous to the conventional definition of integrability in all cases studied so far. \cite{hietarinta2005hirota}.

The Hirota method was used to obtain exact solutions of many important equations such as KdV equation,  modified KdV equation, Sine-Gordon equation, and many other equations \cite{hirota1971exact,hirota1972exact,hirota1972exactB}. This includes also deriving the $N$th-order breather solutions of multi-component nonlocal nonlinear Schrödinger equations \cite{bai2024hirota}, and finding the nondegenerate one-soliton and two-soliton solutions for the generalized coupled higher-order nonlinear Schrödinger equations with variable coefficients \cite{yang2024nondegenerate}.

%%%%%%%%%%%%%%%%%%%%%%%%%%%%%%%%%%%%
Although soliton solutions of different systems, in principle, differ from each other, the soliton solutions of a particular system share some common features. Solutions of a specific system, which share common features are known in this case as a solution family or solution class. Solution classes may be characterized by their stability \cite{malomed2022soliton,zakharov1998optical}, integrability  \cite{gesztesy1992new,malomed2022soliton,khalique2021soliton}, shape preservation  \cite{fedele2002envelope,grimshaw2001two,grimshaw2001two}, asymptotic behavior \cite{gesztesy1992limits,gesztesy1992new,zakharov1998optical}, or energy and momentum \cite{gesztesy1992new,malomed2022soliton,malomed2022soliton,khalique2021soliton,arora2022soliton}. Furthermore, the significance of identifying a solution family is not only in finding the characteristics they have but also in finding more members of that family \cite{gardner1967method,hirota2004direct,gesztesy1992new}.

%%%%%%%%%%%%%%%%%%%%%%%%%%%%%%%%%%%%
To the best of our knowledge, there is no systematic method for finding solution-dependent Hirota bilinear systems of the
nonlinear partial differential equations. Although some researchers implemented the neural network to find bilinear systems and derive new solutions, the solution-dependent bilinear system is not yet seen \cite{zhang2019bilinear, zhang2022bilinear, tuan2024bilinear}. This paper presents a systematic approach to finding the Hirota bilinear system for a given nonlinear differential equation (NLDE). The method uses known exact solutions to the NLDE to identify inter-term cancelations from which a set of equations relating the different terms of the NLDE is obtained. Finally, this set of equations is expressed in terms of the Hirota bilinear system. This is performed using a symbolic code that searches a rather broad range of possible
combinations between the different terms of the NLDE. The method is applied to the NLSE, considering most of its known solutions. Indeed, we found that each class of solutions to this equation is described by a single and different Hirota bilinear system. Finding the bilinear system paves the way for an integrability test, which may lead to new solutions and applications.

The rest of the paper is organized as follows: In Section II, we derive the Hirota bilinear systems for a set of
solutions for NLSE. Then in section III, we derive the bilinear system for the dual nonlinearity NLSE
(flat-top) solution as well as for NLSE with potential. We end in Section~\ref{concsec} with discussion and summary of our main findings.

\section{Hirota bilinear systems for NLSE}
 In this section, bilinear systems are generated for different solution classes, including:
 moving solitons class, and breathers class.
 Each class contains an infinite number of member solutions. For instance, the class of bright solitons
 has single-, two-,
 and multi-bright-soliton solutions.
 The class of breathers  contains: i) Kuznetsov-Ma Breather, ii) Akhmediev Breather, and iii) Peregrine soliton \cite{al2019handbook}.

 Typically, the derivation of Hirota bilinear systems starts with a functional transformation that
 renders the nonlinear differential equation into a form that can be written in the bilinear form. In the
 literature, either rational or logarithmic transformation is used. Here, we use the
 rational transformation, which indeed leads to bilinear systems.

\subsection{Rational transformation}
We consider the rational transformation
 \begin{equation}
 \psi(x,t) = H(x,t)/G(x,t),
 \label{rateq}
 \end{equation}
and substituting it into the  NLSE
 \begin{equation}
     \label{NLSE}
     i\psi_t(x,t) + a_1\psi_{xx}(x,t) + a_2|\psi(x,t)|^2 \psi(x,t) =0.
 \end{equation}
This leads to the rational expression
\begin{equation}
     \label{NLSEHG}
    \sum_{i=1}^7{t_i}=0,
 \end{equation}
where the terms are defined as 
 \begin{equation}
     \label{terms1}
      t_1 = \frac{a_2 |H|^2 H}{|G|^2 G}, \, \, \,
       t_2 = - i \frac{H G_t}{G^2},   \, \, \,   t_3 = i \frac{H_t}{G},  \, \, \,
       t_4 = a_1 \frac{2 H G_x^2}{G^3},  \, \, \,   t_5
       =  - a_1 \frac{2G_x H_x}{G^2}, \, \, \, t_6
       = - a_1 \frac{H G_{xx}}{G^2},  \, \, \, t_7 = a_1 \frac{H_{xx}}{G}.
 \end{equation}

Our method aims at deriving a bilinear form by identifying linear combinations of these terms that cancel each other out, which we refer to as \textit{inter-term cancelations}. Specifically, we search for integer coefficients $j_1, j_2, \dots, j_k \in [-j_{max}, j_{max}]$ such that 
\begin{equation}
q=\sum_{i=1}^k j_i\,t_i\label{terms3}=0,
\end{equation}
where k is the number of terms after the rational transform, which equals 7 in the previous case, and $j_{max}$ can be any number and represents the search domain or the number of iterations in our code. We carefully investigate our search for different values of $j_{max}$ and always get the same results. For example, in the single-bright-soliton solution, we test the search for $j_{max}$ = 4, 6, 8, and 10. However, since the total number of combinations to explore is $(2j_{max}+1)^k$, the process becomes impractical due to exceedingly long processing times. Therefore, we chose $j_{max}$ = 4. To make this search practical,  we proceed in two steps; we first explore cancelations among the terms using a numerical code that tests combinations of integer coefficients $j_i \in [-j_{max}, j_{max}]$. The goal is to identify sets $\{j_i\}$ for which the linear combination $q$ numerically approaches zero (within a small margin of error in our case $10^{-16}$). These are potential solutions where the terms nearly cancel each other out. Once the inter-term cancelations are identified numerically, we verify them symbolically using a Mathematica code. This step confirms the analytic validity of the inter-term cancelations. This combined numerical-symbolic method reduces dramatically the computational time while ensuring correctness. In the following sections, we illustrate this method through several explicit examples.

\subsection{Hirota bilinear systems for specific solutions of NLSE}
We consider here a number of solutions to the NLSE and use them to apply the search procedure described
above. The classes of multi-moving solitons and breathers will be considered. In each case, we find, as
expected, different special Hirota bilinear systems associated with the solutions.\\\\

\paragraph{Class of bright solitons:}
\subparagraph{Single-bright-soliton solution:} The moving single-bright-solution:
\begin{equation}
    \label{BSS}
    \psi(x,t) = A_0 \, \sqrt{\frac{2a_1}{a_2}} \, \, \mathrm{sech}(A_0\, (x-(x_0+t\, v))) {\rm{e}}^{i\,\left(\frac{v}{2a_1}\, (x-x_0)+\frac{4\, A_0^2 \, a_1^2 \, - v^2}{4\,a_1} \,(t-t_0)+ \,\phi_0\right)},
\end{equation}
where $A_0,\,v,\,x_0,\,t_0,\,\phi_0$ are arbitrary real constants corresponding to the soliton's
amplitude, speed, initial position, time start, and initial phase, respectively, are used to define
$H(x,t)$ and $G(x,t)$ as follows
\begin{equation}
    \label{BSSH}
    H(x,t) = 2\,A_0 \, \sqrt{\frac{2\,a_1}{a_2}} \, \, {\rm{e}}^{i\,\left(\frac{v}{2\,a_1}\, (x-x_0)+\frac{4\, A_0^2 \, a_1^2 \, -v^2}{4\,a_1} \, (t-t_0)+ \,\phi_0\right)},
\end{equation}

\begin{equation}
 \label{BSSG}
     G(x,t)= {\rm{e}}^{-A_0 \,(-t\,v +x-x_0)} + \, {\rm{e}}^{A_0\, (-t\,v+x-x_0)}.
 \end{equation}
Now, we use the Mathematica/C++ code to find the inter-term cancelations.  
The inter-term cancelations are found to be
\begin{math}
    \label{BSSeq1}
    q_1 = t_5 + t_2 = 0,
\end{math}
\begin{math}
    \label{BSSeq2}
    q_2 = 2\, t_6 + t_1 + t_4 = 0,
\end{math}
\begin{math}
    \label{BSSeq3}
   q_3 = 2\, t_7 + t_1 + 2\, t_3 + t_4 = 0.
\end{math}
The corresponding bilinear system is obtained as
\begin{eqnarray}
    \label{BSSDa}
    \frac{(q_3 - q_2 + 2\, q_1) \, G^3}{2} = 0 \hspace{0.5cm} &\Longrightarrow &\hspace{0.5cm} G \, (i\, D_t[H.G] +  a_1 \, D^2_x[H.G])=0, \\
    \label{BSSDb}
  q_2 \, G^3 = 0 \hspace{0.5cm} &\Longrightarrow &\hspace{0.5cm}  H \, (a_2 \, |H|^2 \, - \, a_1 \, D^2_x[G.G] )= 0.
\end{eqnarray}

By adding these decoupled bilinear systems, we can get the general bilinear system for the NLSE as follows 
\begin{equation}
\label{GeneralBilinear}
     i\, G \, D_t[H.G] +  a_1 \, G \,  D^2_x[H.G] \, + \, a_2 \, |H|^2 \, H - a_1 \, H \, D^2_x[G.G] =0,
\end{equation}

which is identical to the biliniear system known in the literature \cite{yapicskan8hirota, hietarinta2004introduction}.

\subparagraph{ Two-bright-soliton solution:} The two-soliton solution is derived in 
Ref.~\cite{al2019handbook}. However, we do not present it here for convenience as it is a lengthy expression.
The functions $H(x,t)$ and $G(x,t)$ can be extracted from the solution, as
\begin{eqnarray}
    \label{2BSH}
H(x,t) = - 4\, C  \, \alpha_1 \,  \alpha_2 \, (-\frac{Q_1 + Q_2}{2 \, \alpha_2} + \frac{Q_1 + Q_3}{\alpha_1 + \alpha_2 +i\,  (\nu_1 - \nu_2)}) \nonumber \\ - 4 \, C \,  \alpha_1 \, \alpha_2 \,
(-\frac{Q_3 + Q_4}{2 \, \alpha_1} + \frac{Q_2 + Q_4}{\alpha_1 + \alpha_2 + i\,( \nu_2 - \nu_1)}),
\end{eqnarray}
\begin{equation}
    \label{2BSG}
    G(x,t) = C\, (Q_1 +Q_2)\, (Q_3 + Q_4) - 4 \, (Q_1+ Q_3)\, (Q_2 + Q_4)\, \alpha_1 \, \alpha_2,
\end{equation}
where
\begin{equation}
    \label{Q1}
    Q_1(x,t) = {\rm{e}}^{\frac{(x-x_{02})\, (\alpha_2 - i \, \nu_2)}{\sqrt{2\, a_1}}- i\, \left(\frac{1}{2} \, (t-t_0)\, (\alpha_2 - i \nu_2)^2\right) +\, \phi_{02}},
\end{equation}
\begin{equation}
    \label{Q2}
    Q_2(x,t) ={\rm{e}}^{-\frac{(x-x_{02})\, (\alpha_2 + i\, \nu_2)}{\sqrt{2\, a_1}}- i\, \left(\frac{1}{2} \,(t-t_0)\, (\alpha_2 + i \, \nu_2)^2\right) + \, \phi_{02}},
\end{equation}
\begin{equation}
    \label{Q3}
    Q_3(x,t) ={\rm{e}}^{-\frac{(x-x_{01})\, (\alpha_1 + i\, \nu_1)}{\sqrt{2 a_1}}-i\, (\frac{1}{2}\, (t-t_0)\, (\alpha_1 + i \, \nu_1)^2 + \,\phi_{01}},
\end{equation}
\begin{equation}
    \label{Q4}
    Q_4(x,t) ={\rm{e}}^{\frac{(x-x_{01})\, (\alpha_1 - i \, \nu_1)}{\sqrt{2 a_1}}-i \, (\frac{1}{2} \, (t-t_0) \, (\alpha_1 - i\, \nu_1)^2 + \,\phi_{01}},
\end{equation}
\begin{equation}
 \label{C}
 C = \left(\alpha_1 + \alpha_2 + i\, (\nu_1 - \nu_2)\right)\left( \alpha_1 + \alpha_2 - i\, (\nu_1 -
 \nu_2)\right),
\end{equation}
which leads to the following inter-term cancellations
\begin{math}
    \label{2BSeq1}
    q_1 = 2\, t_5 + t_1 + 2\, t_2 + 4\, t_3 + t_4 = 0,
\end{math}
\begin{math}
    \label{2BSeq2}
    q_2 = 2\, t_6 + t_1 + t_4 = 0,
\end{math}
\begin{math}
    \label{2BSeq3}
    q_3 = t_7 - t_3 = 0,
\end{math}
where $x_{01},\,x_{02},\,t_0,\,\phi_1,\,\phi_2,\,\nu_1,\,\nu_2,\,\alpha_1,\,\alpha_2$ are arbitrary real
constants. The resulting bilinear system is
\begin{eqnarray}
    \label{2BSDa}
    \frac{(q_3 - q_2 + 2\, q_1) \, G^3}{2} = 0 \hspace{0.5cm} &\Longrightarrow &\hspace{0.5cm} G \, (i\, D_t[H.G] +  a_1 \, D^2_x[H.G])=0, \\
    \label{2BSDb}
  q_2 \, G^3 = 0 \hspace{0.5cm} &\Longrightarrow &\hspace{0.5cm}  H \, (a_2 \, |H|^2 \, - \, a_1 \, D^2_x[G.G] )= 0.
\end{eqnarray}

which is identical to the Hirota bilinear system of the single soliton (\ref{BSSDa}) and
(\ref{BSSDb}).
\subparagraph{Finding the Two-bright-soliton solution using the bilinear system:}
Here we derive the two-soliton solution from the bilinear system (\ref{2BSDa}) and (\ref{2BSDb}). Starting by the ansatz solution
\begin{equation}
    \label{ansatz1}
    H(x,t) = \epsilon \,  ({\rm{e}}^{\eta_1(x,t)} + {\rm{e}}^{\eta_2(x,t)}) \, + \,  \epsilon^3 \, (A_1 \,\, {\rm{e}}^{\eta_1(x,t) + \eta_1^*(x,t) + \eta_2(x,t) } +A_2\, \,  {\rm{e}}^{\eta_2(x,t)\,  + \, \eta_2^*(x,t) \, + \, \eta_1(x,t)}),
\end{equation}

\begin{eqnarray}
    \label{ansatz2}
    G(x,t) =  1 \,  +  \,  \epsilon^2 \, (B_1 \,\, {\rm{e}}^{\eta_1(x,t) \, + \, \eta_1^*(x,t)} \, + \, B_2 \,\, {\rm{e}}^{\eta_1(x,t) + \eta_2^*(x,t)} \, + \, B_3 \, \, {\rm{e}}^{\eta_2(x,t) + \eta_2^*(x,t)} \, + \, B_4 \, \, {\rm{e}}^{\eta_1^*(x,t) + \eta_2(x,t)} ) \nonumber \\   +  \, \epsilon^4 B_5 \,  {\rm{e}}^{\eta_1(x,t)+\eta_1^*(x,t) + \eta_2(x,t) + \eta_2^*(x,t)},
\end{eqnarray}
where
\begin{equation}
    \label{eta1}
    \eta_1(x,t) = (C_{11r} +  i\,  C_{11 im} )\, x + (C_{12r} + i \, C_{12 im}) \, t + C_{13r} + i \, C_{13 im},
\end{equation}
\begin{equation}
    \label{eta2}
    \eta_2(x,t) = (C_{21r} +  i\,  C_{21 im} )\, x + (C_{22r} + i \, C_{22 im}) \, t + C_{23r} + i \, C_{23 im},
\end{equation}
\begin{equation}
    \label{eta3}
    \eta_3(x,t) = (C_{31r} +  i\,  C_{31 im} )\, x + (C_{32r} + i \, C_{32 im}) \, t + C_{33r} + i \, C_{33 im},
\end{equation}

\begin{equation}
    \label{A1}
    A_1 = A_{1r} + i \, A_{1 im},
\end{equation}
\begin{equation}
    \label{A2}
    A_2 = A_{2r} + i \, A_{2 im},
\end{equation}
\begin{equation}
    \label{B1}
    B_j = B_{jr} + i \, B_{j im},
\end{equation} 
where j is from 1 to 5.
Then, substituting the above-mentioned equations in the bilinear system, we get

\begin{math}
\label{cons1}
    C_{13r} = C_{13 im} = C_{23r} = C_{23 im} = B_{1 im} = B_{2 im} = B_{3 im} = B_{4 im} =B_{5 im} = A_{1 im} = A_{2 im} = 0,
\end{math}
\\

\begin{math}
    \label{cons2}
    B_{2r} = B_{4r},
\end{math}
\begin{math}
    \label{cons3}
   B_{3r} = \frac{a_2}{8 a_1 C_{21r}^2},
\end{math}
\begin{math}
    \label{cons4}
   B_{1r} = \frac{a_2}{8 a_1 C_{11r}^2},
\end{math}
\begin{math}
    \label{cons5}
B_{4r} = \frac{a_2 C_{22r}^2}{2 a_1 C_{21r}^2 \, (C_{12r} + C_{22r}^2)},
\end{math}
\begin{math}
    \label{cons6}
B_{5r} = \frac{a_2^2 (C_{12r} - C_{22r})^4 \, C_{22r}^2}{
 64 \, a_1^2 \, C_{12r}^2 \, C_{21r}^4 \, (C_{12r} + C{22r})^4)},
\end{math}
\\

\begin{math}
    \label{cons7}
A_{2r} = \frac{a_2 (C_{12r} - C_{22r})^2}{(8 \, a_1 \, C_{21r}^2 \, (C_{12r} + C_{22r})^2)},
\end{math}
\begin{math}
    \label{cons8}
A_{1r} = \frac{a_2\,  (C_{12r} - C_{22r})^2 \, C_{22r}^2)}{8\, a_1 \,  C_{12r}^2 \, C_{21r}^2 \, (C_{12r} + C_{22r})^2},
\end{math}
\begin{math}
    \label{cons9}
C_{22im} = a_1 \, C_{21r}^2 - \frac{C_{22r}^2}{4\, a_1 C_{21r}^2},
\end{math}
\begin{math}
    \label{cons10}
C_{11im} = \frac{-C_{12r}}{2\,  a_1 \, C_{11r}},
\end{math}
\\

\begin{math}
    \label{cons11}
C_{12im} = a_1 \,  C_{11r}^2 - \frac{C_{12r}^2}{4 \, a_1 \, C_{11r}^2},
\end{math}
\begin{math}
    \label{cons12}
C_{21im} = -\frac{C_{22r}}{2\, a_1 \, C_{21r}},
\end{math}
\begin{math}
    \label{cons13}
C_{11r} = \frac{C_{12r} \, C_{21r}}{C_{22r}}.
\end{math}
\\

Substituting back in $G$ and $H$, we obtain the two-soliton solution as the ratio given by

\begin{equation}
    \label{Reverse}
    \psi(x,t) = \frac{H(x,t)}{G(x,t)}.
\end{equation}
We have checked that (\ref{Reverse}) is indeed a solution to (\ref{NLSE}).

This proves that our method works to find the solitons solutions from the bilinear system.

\subparagraph{Three-bright-soliton solution:} The three-bright-soliton is also derived in the literature,
but again, we do not present it here since it is very lengthy \cite{al2019handbook}. Applying our search
procedure leads to the same Hirota bilinear system as for the single- and two-soliton solutions,
(\ref{BSSDa}) and (\ref{BSSDb}). This confirms the fact that solutions within the same class share the same Hirota bilinear system.
\\
\paragraph{Class of dark solitons:}
\subparagraph{Single-dark soliton solution:} The moving single-dark-soliton solution
\begin{equation}
     \label{DSS}
     \psi(x,t)=A_0 \, \sqrt{\frac{-2a_1}{a_2}} \, \, \tanh(A_0 \, (x-(x_0+t v))) \, \, {\rm{e}}^{- \textit{i} \left(\frac{-v}{2a_1} \, (x-x_0)+\frac{8\, A_0^2 \,  a_1^2 - v^2}{4\,a_1} \, (t-t_0)+\, \phi_0\right)},
 \end{equation}
is used to define the $H(x,t)$ and $G(x,t)$ as follows
\begin{equation}
     \label{DSSH}
    H(x,t)=A_0\,  \sqrt{\frac{-2a_1}{a_2}} \, \,  {\rm{e}}^{-i\left(\frac{-v}{2a_1} \,  (x-x_0)+\frac{8 \,  A_0^2 \,  a_1^2 - v^2}{4\, a_1} \, (t-t_0)+\, \phi_0\right)}\, \left ({\rm{e}}^{A_0 \, (x-(x_0 + t v))}- {\rm{e}}^{-A_0 \, (x-(x_0+t v)) }\right),
 \end{equation}

 \begin{equation}
     \label{DSSG}
    G(x,t)={\rm{e}}^{A_0 \, (x-(x_0+t v))} + \, {\rm{e}}^{-A_0\, (x-(x_0+t v))}.
 \end{equation}

The inter-term cancellations turn out to be
\begin{math}
     \label{DSSeq1}
     q_1 = t_1 + t_4 = 0,
 \end{math}
  \begin{math}
     \label{DSSeq2}
     q_2 =2\, t_6 - t_2 - t_5 = 0,
 \end{math}
  \begin{math}
     \label{DSSeq3}
    q_3 = 2 \,t_7 + 3\, t_2 +2\, t_3 + 3\, t_5 = 0,
 \end{math}
 \begin{math}
     \label{DSSeq4}
     q_4 = t_1 + t_2 + t_3 + t_4 + t_5 + t_6 + t_7 = 0.
 \end{math}
The corresponding bilinear system is then given by
\begin{eqnarray}
     \label{DSSDa}
    q_1 \, G^3 = 0  \hspace{0.5cm} &\Longrightarrow &\hspace{0.5cm} H \, (a_2 \, |H|^2 \, + \, 2\, a_1 \, (D_x[G.1])^2) =0, \\
     \label{DSSDb}
    (q_2 \, + \, q_3 \, + \, q_4 \, - \, q1) \, G^3 = 0  \hspace{0.5cm} &\Longrightarrow &\hspace{0.5cm} 
  G \,  (i\,  D_t[H.G] \, + \, a_1\, D^2_x[H.G] - \, 2 \, a_1\,  H \, D^2_x[G.1]) = 0.
\end{eqnarray}

Clearly, this bilinear system differs from the one obtained using the class of bright solitons, namely equations (\ref{BSSDa}) and (\ref{BSSDb}). However, adding equations (\ref{DSSDa}) and (\ref{DSSDb}) yields the general bilinear form given in (\ref{GeneralBilinear}). This confirms that multiple bilinear systems may correspond to the same nonlinear differential equation, depending on the solution class, but they must ultimately lead to the general bilinear form of that equation. \\\\

\paragraph{Class of breathers:}
We now apply our search method to the three breather solutions \cite{al2019handbook}:\\
Kuznetsov-Ma
Breather (KM)
\begin{equation}
    \label{KMB}
    \psi (x,t) = \frac{1}{\sqrt{a_2}}\, \left (\frac{-p^2 \,  \cos(\omega \, (t-t_0))- 2\,i \,p \,\nu \, \sin(\omega\, (t-t_0))}{2\, \cos(\omega\, (t-t_0) - 2 \,\nu \cosh(\frac{p}{\sqrt{2a_1}} \, (x-x_0))} - 1 \right) \, \, {\rm{e}}^{i\, ((t-t_0) + \, \phi_0)},
\end{equation}
Akhmediev Breather
\begin{equation}
    \label{AKH}
    \psi (x,t) = \frac{1}{\sqrt{a_2}}\, \left(\frac{\kappa^2 \, \cosh(\delta \, (t-t_0))+ 2\,i \, \kappa \,  \nu \,  \sinh(\delta \, (t-t_0))}{2 \,  \cosh(\delta \, (t-t_0)) - 2 \, \nu \, \cos(\frac{\kappa}{\sqrt{2 a_1}} \, (x-x_0))} - 1\right) \, \, {\rm{e}}^{i\, ((t-t_0) + \, \phi_0)},
\end{equation}
and the Peregrine soliton
\begin{equation}
    \label{PER}
    \psi(x,t) = \frac{1}{\sqrt{a_2}}  \, \left(\frac{4 + 8 \, i \, (t-t_0)}{1+4\, (t-t_0)^2 + \frac{2}{a_1} \, (x-x_0)^2} - 1\right) \, \,  {\rm{e}}^{i\, ((t-t_0) + \, \phi_0)},
\end{equation}
where, $p = 2 \sqrt{\nu^2 - 1}$, $\omega = p \nu$, $\kappa =2\sqrt{1-\nu^2}$, and $\delta = \kappa \, \nu$. We find that no inter-term cancelations exist except the one which includes all $t_i$-terms. In other
words, all terms in the NLSE couple to each other for the breather to be a solution. Thus, the bilinear system is given by the general bilinear system of NLSE
\begin{equation}
\label{GeneralBilinearBreather}
     i\, G \, D_t[H.G] +  a_1 \, G \,  D^2_x[H.G] \, + \, a_2 \, |H|^2 \, H - a_1 \, H \, D^2_x[G.G] =0.
\end{equation}

\section{Hirota bilinear systems for other versions of the NLSE}
In this section, we derive the Hirota bilinear systems for two different versions of the NLSE,
 namely: A. NLSE with dual nonlinearity, and B. Inhomogeneous NLSE.

\subsection{NLSE with Dual Nonlinearity}
In this section, we consider the Flat-top solution of the NLSE with dual nonlinearity

\begin{equation}
    \label{NLSEDual}
    i \psi_t + a_1 \psi_{xx} + a_2 |\psi|^n \psi + a_3 |\psi|^{m} \psi = 0,
\end{equation}
where $n$, $m$, $a_1$, $a_2$, and $a_3$ are arbitrary real constants. Here we use the case where $m$ = $2n$.
Using the rational transform, this NLSE becomes
\begin{equation}
    \label{NLSEDualHG}
   a_3\,  \frac{|H|^{2n} H}{|G|^{2n} G} + a_2 \frac{ |H|^n H}{|G|^n G} + i \,
   \left(-\frac{H G_t}{G^2} + \frac{H_t}{G}\right) + a_1 \,
   \left(2\,\frac{ H \, G_x^2}{G^3} - 2 \,\frac{G_x \,H_x}{G^2} -
   \frac{H \, G_{xx}}{G^2} + \frac{H_{xx}}{G}\right)=0,
\end{equation}
thus, the terms are
\begin{eqnarray}
    \label{TermsDual}
    t_1 &=&  a_3\,   \frac{|H|^{2n} H}{|G|^{2n} G}, \, \, \, t_2 = a_2  \frac{ |H|^n H}{|G|^n G}, \, \, \,
    t_3 = -i \, \frac{H \, G_t}{G^2}, \, \, \,
    t_4 = i \, \frac{H_t}{G},\, \, \,
    t_5 =  a_1 \, \frac{2 \,H \,G_x^2}{G^3},\, \, \,   \nonumber \\
    t_6 &=&  - a_1 \, \frac{2\, G_x \, H_x}{G^2},\, \, \,
    t_7 =  - a_1 \, \frac{H \, G_{xx}}{G^2},\, \, \,
    t_8 =  a_1 \, \frac{H_{xx}}{G}.
\end{eqnarray}

The flat-top soliton solution
\begin{eqnarray}
    \label{flat}
    \psi(x,t) = \left(\frac{(2\,+n\,) \, A_1 \,\sqrt{\frac{a_2^2 \, (1\,+\, n)}{a_2^2\, (1\,+\,n) + A_1\, a_3 \, (2\, +\, n)^2}}}{a_2 \,  \left(\sqrt{\frac{a_2^2 \, (1\,+\, n)}{a_2^2\, (1\,+\,n) + A_1\, a_3 \, (2\, +\, n)^2}} \,+ \, \cosh
    \left( \sqrt{\frac{A_1}{a_1}}\, n\, \, (x - (x_0 + v \, t)\right)\right)}\right) ^\frac{1}{n}\, \nonumber \\
    \,\times \,{\rm{e}}^{i\, (A_1 (t-t_0) + \, \phi_0)} \, \, {\rm{e}}^{i \, (\frac{v}{2\, a_1} \, (x-x_0) - \frac{v^2}{4 \, a_1} \,(t-t_0) },
 \end{eqnarray}
where $a_1 \,A_1 > 0$, $x_0$,$t_0$, $\phi_0$, and $v$ are arbitrary real constants corresponding to the initial position, initial time, initial phase, and speed respectively, is used to define the $H(x,t)$ and $G(x,t)$ as follows
\begin{equation}
    \label{DNLSEflatH}
   H(x,t) =  (2\,+n\,) \, A_1 \,\sqrt{\frac{a_2^2 \, (1\,+\, n)}{a_2^2\, (1\,+\,n) + A_1\, a_3 \, (2\, +\, n)^2}}\, {\rm{e}}^{i\, (A_1 (t-t_0) + \, \phi_0)} \, \, {\rm{e}}^{i \, (\frac{v}{2\, a_1} \, (x-x_0) - \frac{v^2}{4 \, a_1} \,(t-t_0) },
\end{equation}

\begin{equation}
    \label{DNLSEflatG}
   G(x,t) = \left[{a_2 \,  \left(\sqrt{\frac{a_2^2 \, (1\,+\, n)}{a_2^2\, (1\,+\,n) + A_1\, a_3 \, (2\, +\, n)^2}} \,+ \, \cosh
    \left( \sqrt{\frac{A_1}{a_1}}\, n\, \, (x - (x_0 + v \, t)\right) \right) }\right ] ^\frac{1}{n}.
\end{equation}

We apply the search method for $n$ = 1, 2, and 3.
For $n$ =1, the NLSE becomes
\begin{equation}
    \label{NLSEDualHGN1}
   a_3\,  \frac{|H|^2 H}{|G|^2 G} + a_2 \frac{ |H| H}{|G| G} + i \,
   \left(-\frac{H G_t}{G^2} + \frac{H_t}{G}\right) + a_1 \,
   \left(2\,\frac{ H \, G_x^2}{G^3} - 2 \,\frac{G_x \,H_x}{G^2} -
   \frac{H \, G_{xx}}{G^2} + \frac{H_{xx}}{G}\right)=0,
\end{equation}
the inter-term cancelations are given by

\begin{math}
     \label{DNLSEeq1N1}
     q_1 = t_2+ t_4 = 0,
 \end{math}
  \begin{math}
     \label{DNLSEeq2N1}
     q_2 =t_6 - \frac{1}{3}\, t_1  = 0,
 \end{math}
  \begin{math}
     \label{DNLSEeq3N1}
    q_3 = t_7 - \frac{8}{3}\, t_1 + t_3 + 4 \, t_5 = 0,
 \end{math}
 \begin{math}
     \label{DNLSEeq4N1}
     q_4 = t_8 +4\, t_1 - 3 \,  t_5  = 0.
 \end{math}
The corresponding bilinear system is given by
  \begin{eqnarray}
  \label{DNLSEa}
    (q_3 + \frac{2}{3}\, q_4) \, G^3  = 0  \hspace{0.5cm} &\Longrightarrow &\hspace{0.5cm} \frac{1}{3}\, a_2  \, G \, H\, |H| \, + G \, (i\, D_t[H.G] + \,a_1\, D^2_x[H.G] )= 0, \\
     \label{DNLSEb}
    (q_1 + q_2 + q_3 + q_4)\, G^3 = 0  \hspace{0.5cm} &\Longrightarrow &\hspace{0.5cm} H  \, (\frac{2}{3} \, a_2 \, G\,   |H| \,  + \, a_3\,  |H|^2 \, - \,a_1 \,D^2_x[G.G]) = 0.
\end{eqnarray}

For $n$ =2, the NLSE becomes
\begin{equation}
    \label{NLSEDualHGN2}
   a_3\,  \frac{|H|^4 H}{|G|^4 G} + a_2 \frac{ |H|^2 H}{|G|^2 G} + i \,
   \left(-\frac{H G_t}{G^2} + \frac{H_t}{G}\right) + a_1 \,
   \left(2\,\frac{ H \, G_x^2}{G^3} - 2 \,\frac{G_x \,H_x}{G^2} -
   \frac{H \, G_{xx}}{G^2} + \frac{H_{xx}}{G}\right)=0,
\end{equation}
the inter-term cancelations are given by
\begin{math}
     \label{DNLSEeq1N2}
     q_1 = t_2 + t_4 = 0,
 \end{math}
  \begin{math}
     \label{DNLSEeq2N2}
     q_2 = t_6 - \frac{1}{3} t_1 = 0,
 \end{math}
  \begin{math}
     \label{DNLSEeq3N2}
    q_3 = t_7 -  4\, t_1 +3\, t_5  = 0,
 \end{math}
 \begin{math}
     \label{DNLSEeq4N2}
     q_4 = t_8 + \frac{16}{3} \, t_1 + t_3 -2\,  t_5  = 0.
 \end{math}

The corresponding bilinear system is given by
\begin{eqnarray}
     \label{DNLSEaN2}
    q_1 +  q_2 - \frac{1}{3} \, q_3 = 0  \hspace{0.5cm} &\Longrightarrow &\hspace{0.5cm} - \frac{1}{3}\,  a_3\,  \frac{|H|^4 \, H}{G^2}\,  + \, G\, (i\, D_t[H.G] + \,a_1\, D^2_x[H.G] )= 0, \\
     \label{DNLSEbN2}
    q_4 + \frac{4}{3} \, q_3 = 0  \hspace{0.5cm} &\Longrightarrow &\hspace{0.5cm} H\, ( \, \frac{4}{3} \, a_3\,  \frac{|H|^4 }{G^2}\, + \, a_2 |H|^2  \,   - \,a_1 \, D^2_x[G.G] )= 0.
\end{eqnarray}

For $n$ =3, the NLSE becomes
\begin{equation}
    \label{NLSEDualHGN2}
   a_3\,  \frac{|H|^6 H}{|G|^6 G} + a_2 \frac{ |H|^3 H}{|G|^3 G} + i \,
   \left(-\frac{H G_t}{G^2} + \frac{H_t}{G}\right) + a_1 \,
   \left(2\,\frac{ H \, G_x^2}{G^3} - 2 \,\frac{G_x \,H_x}{G^2} -
   \frac{H \, G_{xx}}{G^2} + \frac{H_{xx}}{G}\right)=0,
\end{equation}
the inter-term cancelations are given by
\begin{equation}
     \label{DNLSEeq1N3}
     q_1 = t_2 + t_4 = 0,
 \end{equation}
  \begin{equation}
     \label{DNLSEeq2N3}
     q_2 =t_1 + t_3+ t_5+ t_6+t_7+t_8.
 \end{equation}

The corresponding bilinear system is given by
\begin{eqnarray}
     \label{DNLSEaN3}
    q_1 + q_2 = 0  \hspace{0.5cm} &\Longrightarrow &\hspace{0.5cm}  a_3\,  \frac{|H|^6 \, H}{G^4} \,  + a_2 \frac{ |H|^3\,  H}{G} \, + G\, (i\, D_t[H.G] + \,a_1\, D^2_x[H.G])  \,- \,\,a_1 \,H\, D^2_x[G.G]= 0.
\end{eqnarray}

\subsection{Inhomogeneous NLSE}
In this section, we apply our method using two solutions for the NLSE with different potentials.
\subsubsection{Bright-Soliton with potential}

Upon using the rational transformation, the inhomogeneous NLSE
\begin{equation}
    \label{NLSEPTBS}
    i \, \psi_t + a_1 \, \psi_{xx} + a_2 \, |\psi|^2 \,\psi + V (x) \,\psi = 0,
\end{equation}
takes the form
\begin{eqnarray}
    \label{fullPT}
   a_2\,\frac{|H|^2 H}{|G|^2 G} \, +\,  \frac{H \, V}{G}
   \,-i\, \frac{H\, G_t}{G^2} \, +\, i\, \frac{H_t}{G} \,+\, 2\, a_2
   \, \frac{H\, G_x^2}{G^3} \,-\, 2 \,a_2\, \frac{G_x\, H_x}{G^2} \,-
   \, a_1 \, \frac{H \, G_{xx}}{G^2} \,+\,a_1 \, \frac{H_{xx}}{G}=0,
\end{eqnarray}
where
\begin{eqnarray}
    \label{termsPT}
   t_1 &=& a_2\,\frac{|H|^2 H}{|G|^2 G}, \, \, \,  t_2 = \frac{H \, V}{G}, \,\,\,  t_3 = -i\, \frac{H\, G_t}{G^2}, \,\,\, t_4 = i\, \frac{H_t}{G},\,\,\, t_5 = 2\, a_2 \, \frac{H\, G_x^2}{G^3},  \nonumber \\ t_6 &=& - 2 \,a_2\, \frac{G_x\, H_x}{G^2}, \,\,\,  t_7 = -a_1 \, \frac{H \, G_{xx}}{G^2}, \,\,\, t_8 = a_1 \, \frac{H_{xx}}{G}.
\end{eqnarray}

The bright-soliton solution and potential
\begin{equation}
    \label{PTBS}
    \psi (x,t) = \sqrt{\frac{V_0}{a_2}} \,\, \mathrm{sech}\left(x \, \sqrt{2\, V_0}\right) \,\,{\rm{e}}^{i\, V_0\, t},
    \end{equation}

    \begin{equation}
    \label{PTBSV}
    V(x) = V_0 \, \mathrm{sech^2}\left(x \,\sqrt{2\, V_0}\right).
\end{equation}

are used to define $H(x,t)$ and $G(x,t)$ as
\begin{equation}
    \label{PTBSH}
   H(x,t) = 2 \, \sqrt{\frac{V_0}{a_2}} \,\, {\rm{e}}^{i \, V_0 \, t},
\end{equation}

\begin{equation}
    \label{PTBSG}
   G(x,t) = {\rm{e}}^{x \, \sqrt{2 V_0}} + \,{\rm{e}}^{- x \, \sqrt{2 \, V_0} }.
\end{equation}

The inter-term cancelations are
\begin{math}
\label{PTBSeq1} q_1 = t_2 - t_1 = 0,
\end{math}
\begin{math}
\label{PTBSeq2} q_2 = t_5 + 2 \, t_1 + 2 \, t_4 = 0,
\end{math}
\begin{math}
\label{PTBSeq3} q_3 = t_7 - t_4 = 0,
\end{math}
\begin{math}
\label{PTBSeq4} t_3 = t_6 = t_8 = 0.
\end{math}

The resulting bilinear system is given by
\begin{eqnarray}
    \label{PTBSDa}
    (2 \, q_3 + q_2) \, G^3 = 0 &\hspace{0.5cm}\Longrightarrow \hspace{0.5cm} & H\, ( 2\, a_2 \, |H|^2 \, -  \, a_1 \, D^2_x[G.G]) = 0,\\
    \label{PTBSDb}
    (q4-q2-2\,q3)\, G^3  = 0 &\hspace{0.5cm}\Longrightarrow \hspace{0.5cm} & H\, (- \, a_2 \, |H|^2 \, + \,G^2 \, V) \, + \,G\, (i\, D_t[H.G]  + a_1 \, D^2_x[H.G]) = 0 .
\end{eqnarray}

\subsubsection{Dark-Soliton with Potential}
Here, the prefactor of the potential in the NLSE will be negative, which changes the sign of $t_2$, and
the rest of the terms will be the same as in the previous section. The NLSE, in this case, is
\begin{equation}
    \label{NLSEPTDS}
   i \, \psi_t + a_1 \, \psi_{xx} + a_2 \, |\psi|^2 \, \psi - V(x) \, \psi = 0. \end{equation}

Using the solution and potential

\begin{equation}
    \label{PTDS}
    \psi (x,t) = A_1 \, \tanh\left(\sqrt{V_0} \, x\right) \, \, {\rm{e}}^{i\, t \, A_1^2 \,  a_2},
\end{equation}

\begin{equation}
    \label{PTDSV}
    V(x) = V_0 \, \, \mathrm{sech^2}\left(\sqrt{V_0} \, x\right),
\end{equation}
the functions $H(x,t)$ and $G(x,t)$ can be written as
\begin{equation}
    \label{PTDSH}
   H(x,t) = A_1 \, \left({\rm{e}}^{x \, \sqrt{V_0}} - \, {\rm{e}}^{-x \, \sqrt{V_0}} \right) \,\,{\rm{e}}^{i\, t\, A_1^2 \, a_2},
\end{equation}
\begin{equation}
    \label{PTDSG}
   G(x,t) ={\rm{e}}^{x \, \sqrt{V_0}} + \,{\rm{e}}^{-x \, \sqrt{V_0}}.
\end{equation}
The inter-term cancelations are
\begin{math}
    \label{PTDSeq1}
    q_1 = t_4 + t_1 + 2\, t_2 = 0,
\end{math}
\begin{math}
    \label{PTDSeq2}
   q_2 =  2 \, t_5 + t_1 = 0,
\end{math}
\begin{math}
    \label{PTDSeq3}
    q_3 = 2 \, t_6 - t_1 - 2 \, t_2 = 0,
\end{math}
\begin{math}
    \label{PTDSeq5}
    q_4 = t_6 - 2 t_7 = 0,
\end{math}
\begin{math}
    \label{PTDSeq4}
    q_5 = t_6 \, + 2 \, t_8 = 0,
\end{math}
\begin{math}
    \label{PTDSeq5}
    q_6 = t_1+ t_2+t_3+t_4+t_5+t_6+t_7+t_8 = 0.
\end{math}

The bilinear system will then be given by
\begin{eqnarray}
    \label{PTDSDa}
    (q_2+q_3 - q_4+q_5)\, G^3 = 0 &\hspace{0.5cm}\Longrightarrow \hspace{0.5cm} &
 H\, (G^2 \, V \, - \, a_1 \, D^2_x[G.G])\, + \, a_1 \, G \, D^2_x[H.G] = 0, \\
    \label{PTDSDb}
    (q_6 -\frac{(q_2+q_3 - q_4+q_5)}{2})\, G^3 = 0 &\hspace{0.5cm}\Longrightarrow \hspace{0.5cm} & H\, (a_2\, |H|^2 \, - \, 2 \, G^2 \, V ) \,  + \, i\, G\,  D_t[H.G]  = 0.
\end{eqnarray}

\section{Discussion and Conclusion}
\label{concsec}
We presented a systematic approach that helps to find Hirota bilinear systems of a given nonlinear partial differential equation. The method generates a bilinear system using known exact solutions to the differential equation. A major finding of the present work is that different bilinear systems may be generated to the same differential equation upon using different exact solutions. Nonetheless, solution members of the same family share a single bilinear system. An example on this case is the family of $N$-soliton solutions, where all members of this family lead to the bilinear system (\ref{BSSDa},\ref{BSSDb}). 

In addition to the fundamental NLSE, we applied our method to the NLSE with dual nonlinearity, and the NLSE with an external potential. In case of NLSE with dual nonlinearity, the family of $N$-flat-top solutions share the bilinear system (\ref{DNLSEa},\ref{DNLSEb}). In case of the inhomogeneous NLSE, namely fundamental NLSE with with a potential, the bilinear system is given by (\ref{PTDSDa},\ref{PTDSDb}). The potentials considered in this case, (\ref{PTBSV}) and (\ref{PTDSV}), resemble the reflectionless P\"oschl-Teller potential, and the resulting inhomgeneous NLSE, (\ref{NLSEPTBS}), admits the exact bright and dark soliton solutions, (\ref{PTBS}) and (\ref{PTDS}), which were used to generate the bilinear systems.

Based on our finding that all members of a solution family correspond to the same bilinear system, the presented method can be used to generate higher order solution members. The lowest order member can be used to derive the bilinear system and then all higher order solution members may be generated using the same bilinear system.

One striking case, though, appears as an example where all terms of the NLSE couple, unlike the case with other solutions where coupling occurs between partial sets of terms. This is the case of breathers. Breathers seem to be special in the sense that they emerge as an exact solution because of coupling among all terms of the NLSE. As a results, we suggest the existence of breathers solutions as a sign on the integrability \cite{hietarinta1997introduction}.
In future work, the method may be applied to a wider range of equations, such as the KdV equation, the Sine-Gordon equation, the Nonlinear Dirac equation, etc. It will be particularly useful to find higher-order solutions of these equations.

Finally, Table (\ref{table1}) summarizes the bilinear system for the classes we found in this paper. 

\section*{Data Availability}
The authors confirm that all data supporting the findings of this study are included within the manuscript.
\clearpage

\begin{table}[]
\caption{Summary}
\begin{adjustbox}{width=\textwidth}
\begin{tabular}{@{}lll@{}}
\toprule
\rowcolor[HTML]{FFFC9E} 
\multicolumn{3}{c}{\cellcolor[HTML]{FFFC9E}Fundamental NLSE  (Eq. \ref{NLSE})} \\ \midrule
\rowcolor[HTML]{EFEFEF} 
\multicolumn{1}{|c|}{\cellcolor[HTML]{EFEFEF}Class} &
  \multicolumn{1}{c|}{\cellcolor[HTML]{EFEFEF}Solution} &
  \multicolumn{1}{c|}{\cellcolor[HTML]{EFEFEF}Bilinear System} \\ \midrule
\multicolumn{1}{|l|}{\cellcolor[HTML]{EFEFEF}} &
  \multicolumn{1}{l|}{\cellcolor[HTML]{EFEFEF}moving one-bright soliton} &
  \multicolumn{1}{l|}{} \\ \cmidrule(lr){2-2}
\multicolumn{1}{|l|}{\cellcolor[HTML]{EFEFEF}} &
  \multicolumn{1}{l|}{\cellcolor[HTML]{EFEFEF}moving two-bright soliton} &
  \multicolumn{1}{l|}{} \\ \cmidrule(lr){2-2}
\multicolumn{1}{|l|}{\multirow{-4}{*}{\cellcolor[HTML]{EFEFEF}Bright Soliton}} &
  \multicolumn{1}{l|}{\cellcolor[HTML]{EFEFEF}moving three-bright soliton} &
  \multicolumn{1}{l|}{\multirow{-3}{*}{\begin{tabular}[c]{@{}l@{}}$G \, (i\, D_t[H.G] +  a_1 \, D^2_x[H.G])=0,$\\ \\ $H \, (a_2 \, |H|^2 \, - \, a_1 \, D^2_x[G.G] )= 0.$\end{tabular}}} \\ \midrule
\multicolumn{1}{|l|}{\cellcolor[HTML]{EFEFEF}Dark Soliton} &
  \multicolumn{1}{l|}{\cellcolor[HTML]{EFEFEF}moving dark soliton} &
  \multicolumn{1}{l|}{\begin{tabular}[c]{@{}l@{}}$ H \, (a_2 \, |H|^2 \, + \, 2\, a_1 \, (D_x[G.1])^2) =0,$\\ \\ $ G \,  (i\,  D_t[H.G] \, + \, a_1\, D^2_x[H.G] - \, 2 \, a_1\,  H \, D^2_x[G.1]) = 0.$\end{tabular}} \\ \midrule
\multicolumn{1}{|l|}{\cellcolor[HTML]{EFEFEF}} &
  \multicolumn{1}{l|}{\cellcolor[HTML]{EFEFEF}Kuznetsov-Ma breather} &
  \multicolumn{1}{l|}{} \\ \cmidrule(lr){2-2}
\multicolumn{1}{|l|}{\cellcolor[HTML]{EFEFEF}} &
  \multicolumn{1}{l|}{\cellcolor[HTML]{EFEFEF}Akhmediev breather} &
  \multicolumn{1}{l|}{} \\ \cmidrule(lr){2-2}
\multicolumn{1}{|l|}{\multirow{-4}{*}{\cellcolor[HTML]{EFEFEF}Breather Solitons}} &
  \multicolumn{1}{l|}{\cellcolor[HTML]{EFEFEF}Peregrine soliton} &
  \multicolumn{1}{l|}{\multirow{-3}{*}{$      i\, G \, D_t[H.G] +  a_1 \, G \,  D^2_x[H.G] \, + \, a_2 \, |H|^2 \, H - a_1 \, H \, D^2_x[G.G] =0.$}} \\ \midrule
\rowcolor[HTML]{FFFC9E} 
\multicolumn{3}{|c|}{\cellcolor[HTML]{FFFC9E}Dual NLSE  (Eq. \ref{NLSEDual})} \\ \midrule
\rowcolor[HTML]{EFEFEF} 
\multicolumn{1}{|c|}{\cellcolor[HTML]{EFEFEF}Solution} &
  \multicolumn{1}{c|}{\cellcolor[HTML]{EFEFEF}Order} &
  \multicolumn{1}{c|}{\cellcolor[HTML]{EFEFEF}Bilinear System} \\ \midrule
\multicolumn{1}{|l|}{\cellcolor[HTML]{EFEFEF}} &
  \multicolumn{1}{l|}{\cellcolor[HTML]{EFEFEF}$n=1$} &
  \multicolumn{1}{l|}{\begin{tabular}[c]{@{}l@{}}$\frac{1}{3}\, a_2  \, G \, H\, |H| \, + G \, (i\, D_t[H.G] + \,a_1\, D^2_x[H.G] )= 0,$\\ \\ $ H  \, (\frac{2}{3} \, a_2 \, G\,   |H| \,  + \, a_3\,  |H|^2 \, - \,a_1 \,D^2_x[G.G]) = 0.$\end{tabular}} \\ \cmidrule(l){2-3} 
\multicolumn{1}{|l|}{\cellcolor[HTML]{EFEFEF}} &
  \multicolumn{1}{l|}{\cellcolor[HTML]{EFEFEF}$n=2$} &
  \multicolumn{1}{l|}{\begin{tabular}[c]{@{}l@{}}$- \frac{1}{3}\,  a_3\,  \frac{|H|^4 \, H}{G^2}\,  + \, G\, (i\, D_t[H.G] + \,a_1\, D^2_x[H.G] )= 0,$\\ \\ $H\, ( \, \frac{4}{3} \, a_3\,  \frac{|H|^4 }{G^2}\, + \, a_2 |H|^2  \,   - \,a_1 \, D^2_x[G.G] )= 0.$\end{tabular}} \\ \cmidrule(l){2-3} 
\multicolumn{1}{|l|}{\multirow{-8.5}{*}{\cellcolor[HTML]{EFEFEF}Flat-Top Soliton}} &
  \multicolumn{1}{l|}{\cellcolor[HTML]{EFEFEF}$n=3$} &
  \multicolumn{1}{l|}{\begin{tabular}[c]{@{}l@{}}$a_3\,  \frac{|H|^6 \, H}{G^4} \,  + a_2 \frac{ |H|^3\,  H}{G} \, + G\, (i\, D_t[H.G] + \,a_1\, D^2_x[H.G])  \,- \,\,a_1 \,H\, D^2_x[G.G]= 0.$ \end{tabular}} \\ \midrule
\rowcolor[HTML]{FFFC9E} 
\multicolumn{3}{|c|}{\cellcolor[HTML]{FFFC9E}NLSE with a potential (Eq. \ref{NLSEPTBS})} \\ \midrule
\rowcolor[HTML]{EFEFEF} 
\multicolumn{1}{|c|}{\cellcolor[HTML]{EFEFEF}Solution} &
  \multicolumn{2}{c|}{\cellcolor[HTML]{EFEFEF}Bilinear system} \\ \midrule
\multicolumn{1}{|l|}{\cellcolor[HTML]{EFEFEF}Bright soliton} &
  \multicolumn{2}{l|}{\begin{tabular}[c]{@{}l@{}}$H\, ( 2\, a_2 \, |H|^2 \, -  \, a_1 \, D^2_x[G.G]) = 0,$\\ \\  $H\, (- \, a_2 \, |H|^2 \, + \,G^2 \, V) \, + \,G\, (i\, D_t[H.G]  + a_1 \, D^2_x[H.G]) = 0 .$\end{tabular}} \\ \midrule
\multicolumn{1}{|l|}{\cellcolor[HTML]{EFEFEF}Dark soliton} &
  \multicolumn{2}{l|}{\begin{tabular}[c]{@{}l@{}}$H\, (G^2 \, V \, - \, a_1 \, D^2_x[G.G])\, + \, a_1 \, G \, D^2_x[H.G] = 0,$\\ \\ $H\, (a_2\, |H|^2 \, - \, 2 \, G^2 \, V ) \,  + \, i\, G\,  D_t[H.G]  = 0.$\end{tabular}} \\ \midrule   
   \label{table1}
\end{tabular}
\end{adjustbox}
\end{table}

\end{document}